\documentclass[sigconf]{acmart}

\usepackage{booktabs}	% 插表格用的宏包
\usepackage{diagbox}    % 插表格用的宏包
\usepackage{multirow}   % 插多行表格用的宏包
\usepackage{verbatim}	% 多行注释用的宏包
\usepackage{array}    % 表头自定义对齐方式
\usepackage{balance} % 最后一页对齐
\usepackage{amsmath,amssymb,amsfonts,graphicx}
\usepackage{epstopdf}
\usepackage{stfloats} % 图表放置当前页底部

\usepackage[noend]{algpseudocode}
\usepackage[ruled,linesnumbered]{algorithm2e}

\AtBeginDocument{%
  \providecommand\BibTeX{{%
    \normalfont B\kern-0.5em{\scshape i\kern-0.25em b}\kern-0.8em\TeX}}}

\copyrightyear{2023}
\acmYear{2023}
\setcopyright{acmlicensed}\acmConference[WWW '23 Companion]{Companion Proceedings of the ACM Web Conference 2023}{April 30-May 4, 2023}{Austin, TX, USA}
\acmBooktitle{Companion Proceedings of the ACM Web Conference 2023 (WWW '23 Companion), April 30-May 4, 2023, Austin, TX, USA}
\acmPrice{15.00}
\acmDOI{10.1145/3543873.3587583}
\acmISBN{978-1-4503-9419-2/23/04}

\begin{document}

\title{Web 3.0: The Future of Internet}

%% Frequent Itemset Mining


\author{Wensheng Gan}
\authornote{Also with Pazhou Lab, Guangzhou 510330, China}
\affiliation{%
	\institution{Jinan University} %College of Cyber Security
	\city{Guangzhou}
	\country{China}
}
\email{wsgan001@gmail.com}

\author{Zhenqiang Ye}
\affiliation{%
	\institution{Jinan University} 
	\city{Guangzhou}
	\country{China}
}
\email{yzq66f@gmail.com}

\author{Shicheng Wan}
\authornote{Corresponding author: scwan1998@gmail.com}
\affiliation{%
	\institution{Guangdong University of Technology} 
	\city{Guangzhou}
	\country{China}
}
\email{scwan1998@gmail.com}

\author{Philip S. Yu}
\affiliation{%
	\institution{University of Illinois at Chicago}
	\city{Chicago}
	\country{USA}
}
\email{psyu@uic.edu}

\renewcommand{\shortauthors}{Gan et al.}

\begin{abstract}
   With the rapid growth of the Internet, human daily life has become deeply bound to the Internet. To take advantage of massive amounts of data and information on the internet, the Web architecture is continuously being reinvented and upgraded. From the static informative characteristics of Web 1.0 to the dynamic interactive features of Web 2.0, scholars and engineers have worked hard to make the internet world more open, inclusive, and equal. Indeed, the next generation of Web evolution (i.e., Web 3.0) is already coming and shaping our lives. Web 3.0 is a decentralized Web architecture that is more intelligent and safer than before. The risks and ruin posed by monopolists or criminals will be greatly reduced by a complete reconstruction of the Internet and IT infrastructure. In a word, Web 3.0 is capable of addressing web data ownership according to distributed technology. It will optimize the internet world from the perspectives of economy, culture, and technology. Then it promotes novel content production methods, organizational structures, and economic forms. However, Web 3.0 is not mature and is now being disputed. Herein, this paper presents a comprehensive survey of Web 3.0, with a focus on current technologies, challenges, opportunities, and outlook. This article first introduces a brief overview of the history of World Wide Web as well as several differences among Web 1.0, Web 2.0, Web 3.0, and Web3. Then, some technical implementations of Web 3.0 are illustrated in detail. We discuss the revolution and benefits that Web 3.0 brings. Finally, we explore several challenges and issues in this promising area.
\end{abstract}




\maketitle

\section{Introduction} \label{sec:introduction}
% Table: Webs introduction
The history of the development of the World Wide Web consists of four phases, i.e., Web 1.0, Web 2.0, Web 3.0, and Web3. Thanks to scientific and technological innovation, we have experienced the Web 1.0 era and live in a period of Web 2.0, Web 3.0, and Web3 coexistence. The innovator of the World Wide Web, Timothy John Berners-Lee\footnote{\url{https://en.wikipedia.org/wiki/Tim_Berners-Lee}} proposed a distributed hypertext system, called the Web (or Web 1.0) \cite{berners1989information}. He also designed and built the first web browser\footnote{WorldWideWeb: \url{https://worldwideweb.cern.ch/worldwideweb/}} and published the first website\footnote{\url{http://info.cern.ch/} or \url{https://worldwideweb.cern.ch/browser/}}. The initial Web is a linked information system, which is based on graph and link organization mode. A significant feature of Web 1.0 applications is static pages. Visitors are permitted to perform a few simple operations, such as reading and clicking. It is so monotonous that few people were interested. After that, Web 1.0 applications continued to evolve to be more versatile and easier to use. The second-generation Web, called Web 2.0, was proposed in a brain-stream forum \cite{san2007under}. Compared to Web 1.0, users are no longer just reading or downloading content from static websites. They are capable of writing or uploading various creations on the internet. The interaction is vital for Web 2.0 architecture. A lot of novel technologies (e.g., asynchronous JavaScript and XML, Cascading Style Sheets, the document object model, and JavaScript object notation) make users enjoy rich experiences \cite{o2007web}. Until now, Web 2.0 inspired many young people's creative enthusiasm and encouraged them to participate. Social media platforms (e.g., Facebook, WeChat, TikTok, and Twitter), video and music websites (e.g., YouTube, BiliBili, and Spotify), and e-business platforms (e.g., Amazon, Tmall, eBay, and Walmart) are relatively mature and full of all aspects of our lives in the past decades \cite{choudhury2014world}.

However, on the one hand, whether users are voluntary or not, their application data belongs to the corresponding Web 2.0 platforms; on the other hand, these platforms collect users' data as much as possible and then maximize their revenue \cite{alabdulwahhab2018web}. It should be pointed out that users generally do not know how and for what their information will be used. Web 3.0 \cite{berners1998semantic,geeks2021how} provides a more transparent architecture (i.e., decentralized). In the view of Timothy John Berners-Lee, Web 3.0 aims to create a more intelligent web, which emphasizes machine understanding of human semantic expression \cite{berners1998semantic,berners2001semantic,shadbolt2006semantic}. Later, Ethereum co-founder Gavin Wood thinks that centralized services cause a lot of corporate monopolies, and thus the next Web flood will completely change the status of Web 2.0 \cite{wood2014ethereum}. He believes that the next generation of the Web will be an identity-based pseudonymous low-level messaging system. In order to make it distinct from traditional Web 3.0, he renamed it Web3. Web3 architecture achieves decentralization via blockchain technologies, whereas Web 3.0 may not require blockchain. Today, Web 3.0 is a broad but borderless concept. It has integrated powerful and large-scale Web applications. We suppose Web 3.0 is a powerful, generic, and measurable architecture.

% Table: differences among Webs
\begin{table*}[ht]
    \centering
    \caption{Differences among different stages of webs}
    \label{tab:diffWebs}
    \begin{tabular}{| m{1cm}<{\centering} | m{2cm}<{\centering} | m{3.6cm}<{\centering} | m{3.4cm}<{\centering} | m{3.4cm}<{\centering} |}
        \hline
        \textbf{Web}  & \textbf{Architecture} & \textbf{Representative products} & \textbf{Characteristics} & \textbf{Benefit distribution}  \\ \hline
			
        Web 1.0 & centralized & Yahoo, Sina, Netscape & host-generated content, host-generated authority & platform monopoly \\ \hline
			
        Web 2.0 & centralized & Baidu, Google, Facebook & user-generated content, host-generated authority & profit-sharing (platforms and netizens) \\ \hline
			
        Web 3.0 & distributed, decentralized & Tor, Twine & user-generated content, user-generated authority & peer-to-peer \\ \hline
			
        Web3 & distributed, decentralized & Ethereum, Binance & user-generated content, user-generated authority & smart contract \\
	\hline
    \end{tabular}
\end{table*}

Table \ref{tab:diffWebs} summarizes the differences between four types of Webs. Web 1.0 is the informational internet. It only offers a reading experience for users; there is no interaction or dynamic content. Web 2.0 is synonymous with identity and a centralized Web. Users become content creators and are willing to communicate with others through Internet tools. However, it is hard to break down the information blockade between platforms. Web 3.0 and Web3 are both user-generated content and user-generated authority. That is, users can determine what and how much information other people and platforms can view. This allows users to truly own their data.

%\textbf{Research gap}: 
There are some reviews of literature related to Web 3.0 \cite{rudman2015web,o2021too,weaver2021web3,gavin2018why,liu2021make,korpal2022decentralization}. Most of them had not clearly distinguished between Web 3.0 and Web3. For instance, because blockchain is known for implementing a new organization governance model (i.e., decentralization), it is easy to take it for granted that blockchain technology is the most suitable tool within Web 3.0 architecture. However, current blockchain technology is not mature, and its financial hype is concerning (see \cite{chohan2022cryptocurrencies}). Though Web 3.0 is a buzzword, most people are still unclear about it, especially its definition. Web 3.0 will facilitate a worldwide data reform, which may trigger opportunities and risks. There is no doubt that providing a detailed illustration (what, how, and when) is valuable.

%\textbf{Contributions}: To the best of our knowledge, this is the first up-to-date review on Web 3.0.
To fill this gap, this paper aims to conduct a systematic literature review of Web 3.0. The key contributions of the article are fourfold.

\begin{itemize}
    \item We elaborate on the evolution history of the World Wide Web, which reveals that there have been unprecedented advances in the pursuit of democracy within the digital world.

    \item We introduce some vital Web 3.0 technologies, from data storage to information analysis and usage. Additionally, We identify the key difference between Web 3.0 and Web3.
    
    \item In particular, we provide a detailed survey of the Web revolution and the benefits in every aspect of our lives (including business, culture, Metaverse, and AI-generated content, etc.) that Web 3.0 brings. 
    
    \item Finally, we highlight and discuss some key challenges and future work based on our review. We also make some recommendations for Web 3.0 governance and development.
\end{itemize}

\textbf{Organization}: This article summarizes recent Web 3.0 advancements. In Section \ref{sec:technical}, we introduce several vital technologies within the Web 3.0 architecture. The revolution and benefits of Web 3.0 are then briefly introduced in Section \ref{sec:revolution}. In Section \ref{sec:challenges}, we discuss in detail the existing or emerging challenges and issues of Web 3.0. We also discuss the difference between Web 3.0 and Web3 in Section \ref{sec:discussion}. Finally, in Section \ref{sec:conclusion}, we conclude this article with discussions and potential future work.

\section{Technical Implementation} \label{sec:technical}

Despite the lack of a unified definition of Web 3.0, major features such as decentralization, privacy protection, human centricity, and intelligence are widely accepted \cite{silva2008web}. In the Web 3.0 world, machines will better understand human behavior, which will provide more intelligent services. In this section, we introduce some technologies that might become cornerstones of Web 3.0.

% 语义网 以及知识图谱
\subsection{Semantic Web}

The semantic web \cite{berners1998semantic}, a prototype of Web 3.0, is an information-data web that aims to concatenate all the data in the virtual world. With the rapid development of science and technology, proposals to manage the abundance of web data (e.g., data sharing, integration, reuse, and mining) are one of the major obstacles \cite{shadbolt2006semantic, halfpenny2009special}. Resource Description Framework (RDF) \cite{miller1998introduction} is a syntax-neutral data model (i.e., \textit{Subject}, \textit{Predicate} and \textit{Object}). RDF records the relationship between elements \textit{Subject} (e.g., links) and \textit{Object} (e.g., resources) and describes the features of web resources. It mainly provides an infrastructure for various applications of metadata and exchanges metadata between applications on the Web, which promotes the automatic processing of network resources \cite{choudhury2014world}. Subsequently, the Web Ontology Language (OWL) \cite{mcguinness2004owl} was proposed to improve the comprehensibility of web content for machines and play a part in semantic web activity. It is a family of knowledge representation languages for authoring ontologies. Ontologies resemble class hierarchies in object-oriented programming, and the core idea of OWL is to represent the ontology explicitly and efficiently \cite{hitzler2021review}. The OWL is used to make network resources more accessible for automated processes by adding resource information that describes or provides web content. Besides, knowledge graphs (KG) \cite{singhal2012introducing, bonatti2019knowledge} may be the next direction for knowledge representation on the semantic web. A knowledge graph consists of a set of interconnected typed entities and their attributes \cite{atzori2020special, bernstein2016new}. According to the study \cite{fensel2020introduction}, there are four main steps for KG generation: 1) knowledge creation; 2) knowledge hosting; 3) knowledge curation; and 4) knowledge deployment. KG may be the most possible way to achieve the ``Internet of Behaviors'' (IoB) blueprint \cite{sun2023internet}, which can build connections between people and things, or things and things.

% 人工智能
\subsection{Artificial Intelligence}

Benefiting from the improvement of computing power and big data technologies \cite{gan2017data}, AI has ushered in a period of vigorous development. AI is becoming a part of our daily lives as more domains deploy AI applications \cite{haenlein2019brief, hailemariam2020empirical}. We can provide numerous datasets and use AI training models for solving problems such as image recognition, information extraction, and automatic speech recognition. In the Web 3.0 era, massive amounts of data will be generated every day from device perception, content services, and intelligent life. AI helps machines realize the ``perception-decision-behavior-feedback'' closed-loop workflow, and thus improve the user experience. Moreover, since the integration of computing and storage breaks the bottleneck of AI computing power, the development of IoT collaborative perception and 5G communication technologies will realize the collaboration between multiple agents, which can meet people's needs for real-time perception and decision-making. Many other fields have made great progress by enabling AI. For example, autonomous driving \cite{ding2020large}, according to the in-depth integration of IoT and AI, offers the best route planning and control for vehicles. Market forecasting and risk management in financial markets, medical assistance in the health industry, recommendation systems, unmanned retail in the retail industry, voiceprint payment, face scanning in payment systems, and voice in the smart home are all examples of how technology is changing our lives \cite{zhang2021study}. All of the above cases illustrate how AI makes Web 3.0 more intelligent and user-friendly. However, because AI products have a great impact on our lives, fairness, and non-discrimination (including objective and subjective) in the development of AI will be particularly important. For example, the usage of big data is unethical and malicious behavior by companies toward their customers. In some cases, AI products serve some groups but ignore the requirements of specific groups (e.g., the elderly and the disabled). In the Web 3.0 era, data ownership belongs to users because they generate new data every day. These data may be meaningless to users, but companies can profit from a variety of data using AI technologies, such as user profiles and personalized advertising. The definitions of fairness are distinct in different historical periods and even in different ideologies. Fortunately, AI technologies can improve fairness and transform it into a global and comprehensive understanding, which provides a powerful guide to achieving fairness. Moreover, with the development of technology (e.g., federated learning, trusted computing, the Internet of Things, Internet of Behaviors, and encryption), most negative effects that technology brings will be eliminated in most cases \cite{chen2022federated, chen2022metaverse}.

% 区块链, 语义区块链，知识图谱区块链
\subsection{Blockchain}

The emergence of an embryonic blockchain was taking shape from the 1980s to the 1990s and was officially released in 2008 \cite{nakamoto2008bitcoin, yaga2019blockchain}. Many experts, scholars, and capitalists are interested in the potential of blockchain technology because of its decentralization, trustlessness, autonomy, anonymity, tamper-proofing, and auditability \cite{zheng2018blockchain}. One of the most famous successful cases is Ethereum \cite{buterin2014next}. Ethereum provides a built-in Turing-complete programming language, which can help developers code smart contracts and build their own decentralized applications. Ethereum yellow paper (``a canonical version'') \cite{wood2014ethereum} provided a quasi-Turing-complete machine called the Ethereum Virtual Machine. To protect smart contracts from malicious attacks, the Ethereum Virtual Machine provides a sandbox execution environment. Arguably, the proposed novel underlying technologies, such as token systems, identity and reputation systems, decentralized file storage, and decentralized autonomous organizations within the blockchain, can help fight the monopoly posed by giant tech companies. Some metrics are proposed to evaluate the decentralization of blockchain. For example, the studies \cite{gencer2018decentralization,chu2018curses, kim2021impossible} illustrate several related metrics. Croman \textit{et al.} \cite{croman2016scaling} proposed a more intuitive method. They believe that the more active addresses on the blockchain, the better the blockchain's decentralization.

Moreover, semantic blockchain and knowledge-based blockchain may be the next technologies among the most widely accepted in Web3 (not Web 3.0) \cite{ruta2017semantic}. The internet service within semantic and knowledge-based blockchains not only has decentralized and trustless features but also takes advanced advantage of artificial intelligence. What's more, the ``impossible trinity problem'' (that is, decentralization, privacy, and scalability) seems unsolvable for a long time. This case will urge researchers and developers to explore other paths for decentralization solutions thereafter.

% 去中心化存储 IPFS 
\subsection{Decentralized Storage}

Before starting this subsection, we have to point out that decentralized storage is not a necessary part of Web 3.0 architecture, but decentralization will be more secure and reliable than centralization. There has long been some sort of unspoken agreement between users and Web 2.0 applications. That is, users' data belongs to platforms, and users just use the services provided by those platforms. Moreover, ``data island'' between different platforms also brings many barriers, such as data migration and data synchronization. In the meanwhile, while users realize their data autonomy, lowering the data storage cost and finding a suitable benefit distribution are urgent problems to be solved. In order to implement decentralized data storage, researchers made the following contributions:

 \textbf{IPFS:} The Interplanetary File System (IPFS) \cite{case2015why,benet2014ipfs} is a peer-to-peer distributed file system, which may replace HTTP\footnote{\url{https://en.wikipedia.org/wiki/Hypertext_Transfer_Protocol}} in the future. It splits files into several blobs (an addressable unit of data with no links, and its size will not be larger than 256 KB). These blobs are organized by the file object ``list'' or alternatively ``tree''. The blob has a hash fingerprint, which is recorded in the distributed hash table according to a value-key structure ($<$hash fingerprint, location node$>$). IPFS adopts the Merkel directed acyclic graph to locate content and deduplication. Moreover, IPFS provides a version control function like Git for helping user management. In order to adapt to users' reading habits, the adopted Inter-Planetary Naming Service maps the URLs to a series of hashes in the IPFS system. More importantly, the sharing strategy BitSwap prevents freeloaders from exploiting and degrading the exchange \cite{manoj2021peer}.
    
 \textbf{CephFS:} Ceph file system (CephFS) \cite{weil2006ceph} is known for being open-source, distributed, high-performance, highly available, and scalable. The Reliable Autonomic Distributed Object Store (RADOS) clusters are its core part. The cluster consists mainly of object storage devices (OSDs), monitors, and clients. OSD is a storage node in a cluster used for data storage and maintenance. The metadata server cluster can expand or contract, and it can rebalance the file system dynamically to distribute data evenly among cluster hosts. Most importantly, RADOS does not rely on a single central control component. Heavy loads all can be dynamically distributed within the cluster. Thus, this makes sure that the physical decentralization and high scalability of CephFS.

There are still many decentralized storage projects out there. For example, OpenStack Object Storage (Swift) \cite{rosado2014overview}, uses consistent hashing technology to evenly distribute objects to each virtual node and uses the ring structure to store the mapped physical address of virtual nodes. Finally, high availability and infinite horizontal expansion ability are achieved by using the event-driven consistency model. Unlike the initial decentralized storage scheme, the current hot decentralized storage scheme considers the user to be the subject of participating in the data storage link. IPFS introduces the BitSwap protocol to reward users for providing resource storage and download services, thus ensuring the possibility of users benefiting from data. Existing projects like Filecoin and ARweave are based on IPFS, which can be considered the incentive layer of IPFS. There are other star projects, like Swarm for the Ethereum foundation. The use of blockchain technology in conjunction with a reward mechanism is becoming more common.

% 边缘计算或者算力网络：云计算，5G，边缘计算，结合联邦学习, 结合物联网
\subsection{Edge Computing}

Because of the COVID-19 epidemic, more and more of our daily activities are happening online, which has led to a sharp rise in the amount of data and network traffic around the world. The total amount of data in the world in 2020 was already 59 ZB. Under the impact of such a huge data stream, all data processing is placed on the remote server, which will be a great challenge to the network, computing resources, and so on. Although some data processing problems can be alleviated through cloud computing, the proportion of valuable data in massive data sets is very low. Therefore, it is necessary to process the data at the beginning of its generation and then analyze it through the cloud. In order to solve this problem, edge computing and fog computing were proposed \cite{zhou2019edge,cao2020overview,hailemariam2020empirical}. Some tasks are processed through edge devices, in this way, the pressure of cloud computing can be shared. The main difference between edge computing and fog computing is that edge computing allows the end devices to process data by themselves \cite{varghese2016challenges}. The main idea of edge computing is to migrate core functions like computing, storage, and decision-making closer to the edge devices that generate the data. As a result, the edge computing model eliminates the need to upload data to a cloud computing platform for storage and processing. Moreover, as edge computing is closer to the data source, it has fast data processing and analysis and low cost, low energy consumption, and low bandwidth. Although edge computing has a certain improvement in the security aspect when compared with cloud computing since it can avoid the risks during the network transmission process, the edge device obtains first-hand data, which has a large amount of sensitive private information. Due to the lack of effective encryption or desensitization methods for data, when an emerging hacker attacks the edge node, critical information such as household personnel consumption, personnel health information in the electronic medical system, and road incident vehicle information will be leaked. Federated learning (FL) \cite{wang2019adaptive} trains AI models by coordinating multiple remote devices and not directly exposing users' data, which enhances the privacy of data. The combination of edge computing and FL will provide a more secure environment for edge nodes and protect the information security of users. In general, edge computing can effectively explore the potential of edge devices and provide users with more high-quality services in the era of big data.

\section{Revolution and Advantages} \label{sec:revolution}

% 商业模式
\subsection{Application and Business}

\textbf{Web 1.0}. In the early years of Web 1.0, the proliferation of the birth of the World Wide Web spurred a growing need for knowledge sharing that cannot be met solely by traditional information transmission methods. Many companies, like Sina, Yahoo, Google, and Baidu, have promoted their own products. As shown in Figure \ref{tab:products_web_1.0}, these products can be roughly divided into two categories. The first one is the web portal represented by Sina and Yahoo. In view of knowledge creation, these web portals aim to digitalize human knowledge from offline to earn profit online through clicks. The second is represented by the search engines of Google Search and Baidu. They are used to answering users' questions as accurately as possible. The search engine does not actively share content with users. It seems like a library on the Internet and automatically collects or categorizes various types of information. Through continuous development, Google Search has been one of the most intelligent and powerful search engines in the world. As mentioned earlier, both the two categories of web products are host-generated content and authority. On the one hand, users can only read what websites have to offer; on the other hand, Web 1.0 enabled numerous people to learn about and engage with the Internet.

\textbf{Web 2.0}. With the rapid growth of the number of internet users in the world, the occurrence of Web 2.0 has attracted considerable attention. Unlike Web 1.0, the new Web lets users create content on many different online platforms. In other words, a significant difference between Web 1.0 and Web 2.0 is that the latter is user-generated content rather than host-generated content. According to Web 2.0 products, both users and platforms can earn profits. The current web products have more vertical subdivisions than in the Web 1.0 era. Figure \ref{tab:products_web_2.0} lists some major applications in our daily lives. Herein, we plan to introduce some of them to reveal the success of Web 2.0. WeChat is very popular in Asian regions. It replaces the traditional text message service of mobile operators to a certain degree because of its novel functions (e.g., video calls and group chats). Twitter is one of the most famous social network applications now. With unique characteristics, such as character limits and photo sharing, it provides text message services over the internet for users. Twitter is also an information-sharing platform and forms many grid communities. People around the world can share their daily lives or interact with others by posting a tweet. The i Operating System (iOS) is a mobile device system developed by Apple Inc. In the Web 2.0 era, the mobile phone totally changes the lifestyle of humans. iOS and its competitors (Android\footnote{\url{https://en.wikipedia.org/wiki/Android_(operating_system)}}) are vehicles that run applications from Web 2.0. Due to the breakthrough in hardware research, the applications of Web 2.0 are more powerful and practical than the websites of Web 1.0. These applications are convenient for human activities, and most of us cannot imagine how to live without the Internet.

\textbf{Web 3.0}.  As reported in some white papers \cite{assessingiapp2017,theosterman2022,mobilecaict2021}, the rapid growth of the current digital economy, especially after the COVID-19 outbreak, has led to a lot of monopolist giants (e.g., Facebook, Google, Tencent, and Amazon). Now we face the challenge of extracting the world from the jaws of online platform monopolies. The applications within Web 2.0 are more likely to provide services rather than products. This causes the traditional concept of data ownership to become blurry. Besides, most Web 2.0 companies are profiting from users' private information. The study \cite{susanto2021digital} points out that digital ethics and privacy issues within the internet need more attention. Web 3.0 is engendering a new global digital economy. It creates new business models and markets to go with them, and it busts platform monopolies. Web 3.0 applications are deployed on decentralized networks such as blockchain platforms or related distributed systems hosted by many peer-to-peer servers. They are designed based on different scenarios. In order to concretely study the difference between applications of Web 2.0 and Web 3.0, we list mainstream applications within the same classification in Figures \ref{tab:products_web_2.0} and \ref{tab:products_web_3.0}, respectively. Status is a secure messaging application that provides private communication. It uses peer-to-peer technology to prevent any third party from controlling users' communication data. This is totally different from WeChat. On account of our data being stored on the platform, there are potential attack risks like data trawling, censorship, and propaganda \cite{steinberg2022cyber}. The pseudo-anonymous account generation allows users to selectively reveal themselves to the world. Status, in particular, is an entirely open-source project. This is a common characteristic of most Web 3.0 applications, which can ensure the application is not malicious. Steemit \cite{chohan2018concept} is a rising star among the many Web 3.0 social network applications. It is a blockchain-based platform that aims to return data power to its users rather than centralizing control by traditional social media companies. Steemit also uses the eponymous cryptocurrency STEEM\footnote{\url{https://observer.com/2016/09/steem-tsu-social-networks-spam/}} to reward users for their content. Electro-Optical Systems (EOS) is a blockchain operating system that provides the core functionality for businesses to build blockchain applications. EOS is similar to the Windows platform, and the system architecture is EOS.IO\footnote{\url{https://en.wikipedia.org/wiki/EOS.IO}}. Compared to Android and iOS, EOS runs on a public chain. The user only needs a browser that can link to EOS within a mobile phone, iPad, or computer device, which is truly cross-platform.

% Table: Web 3.0 case

\subsection{Culture and Artwork}

Indeed, the interaction between culture and society is common. What's more, it should be pointed out that: on the one hand, cultural production may not generate social activities; on the other hand, social activities may not create new cultural products. Cultural activities can be roughly divided into two parts: cultural creation and cultural communication. The process of culture creation within Web 3.0 architecture is more transparent than ever before. It allows other people to freely engage in co-creation. Because each person's contribution is clear, it is simple to distribute the benefit and copyright. At the same time, every node can make peer-to-peer transactions since Web 3.0 is a decentralized network. Thus, there is no need for a third-party agent or platform to help creators sell their products. This case not only increases creators' earnings but also provides them with more options. Furthermore, Web 3.0 combines various types of information based on the needs of the users and then provides personalized recommendation services. Since the profit belongs to the creators themselves, they have more incentive to promote their works. Cultural heritage or productions are problematic because they are non-renewable, fragile, and expensive to maintain. They can be digitized by massive sensors, according to Web 3.0 technologies. The methodology of digital humanities brings about fundamental changes in cultural heritage. Through digital methods, cultural heritage can be better protected, and cultural information can be linked to the spatio-temporal framework. The transformation of time and space can activate more users' participation enthusiasm and revitalize the vitality of cultural heritage.

\subsection{User Experience}

The forthcoming wave of Web 3.0 is promoting technical integration across different domains. It may be the most anticipated event during the 21st century. We plan to roughly discuss the user experience in terms of compatibility, permissionlessness, and availability.

\textbf{Compatibility:} As we mentioned before, different Web 2.0 applications (either homogeneously or heterogeneously) often have data segregation problems when transferring data across products, such as a uniform format, inconsistent data, distinct coding mechanisms, etc. However, Web 3.0 provides many standardized APIs that solve this problem to a certain extent. Applications and services are no longer limited by a single ecosystem (e.g., the Ethereum ecosystem and Bitcoin). Web 3.0 is a new integrated ecosystem that is compatible with Web 2.0 to a certain extent. For example, Ceramic\footnote{Ceramic network: \url{https://ceramic.network/}} tries to build applications with composable Web 3.0 data and enable reusable data for multiple scenarios. The unique digital wallet within Web 3.0 is adapted to store user internet data, which brings a number of benefits. For example, the encryption technologies can ensure the privacy and safety of a digital wallet, and the third party can ask for data reading permission from the user. This solves the problem of data silos, which belong to Web 2.0, and users of Web 3.0 do not have to worry about loss of data issues from replacing devices or platforms anymore.
    
\textbf{Permissionless:} Meanwhile, users' application data no longer belongs to online platforms or governments. Users can fully control assets and metadata, and optionally release them to service providers according to their personal preferences. For instance, Timothy John Berners-Lee proposed a novel Web data protocol, named Social Linked Data\footnote{\url{https://solidproject.org/}}, to constrain web infringing activities. He also designed the POD, which can be established on a personal server or hosted by a third-party platform, to store user data. Other applications can only request the users' authorization to obtain the permitted information. The user application data (e.g., account, password, browsing history, bookmarks, and related items) will only be recorded on the POD. In other words, the user data is no longer bounded to any platform, but the platform needs to request that the user read the needed data. In this case, users can directly interact with others who they are not familiar with or without the need for a trusted third-party platform. The only thing that needs to be provided is the user's private key or other identifiable proof.
    
\textbf{High availability:} Web 3.0 is an open and free world where all web data is stored in public community networks. What's more, a huge number of data put forward higher criteria for the new network architecture: a high fault tolerance rate and low fault probability. This means that users are capable of using data normally in abnormal environments and scenarios. In addition, easy-to-use is one of the highlights of Web 3.0 applications and software. The design of application or software enables users to focus on their perception, their own tasks, and their operations according to their own course of action. They are not required to be distracted by searching for the human-machine interface's menu or understanding the structure of software, the human-machine interface, and the meaning of the icons \cite{pan2006virtual,flavian2019impact}. They also do not have to consider how to convert their tasks into the input mode and steps of the machine. Because the virtual world is more integrated with the real world, users are no longer limited to the previous input devices, such as keyboards or microphones. A look or a simple action (e.g., raising hands and nodding) are input instructions for machines.

\subsection{Metaverse}

The word ``meta'' means beginning, important, and consummation. The Metaverse is a mixture of virtual and real worlds. The Web 3.0 architecture provides lower-level support for building Metaverse applications. Immersive interactive technology (e.g., VR, AR, and MR) can create a more attractive digital living space for users. The Metaverse will significantly change the following domains: 1) Education and training. A virtual educational environment facilitates educators' ability to teach students. Besides, immersive, interactive learning environments let students better understand knowledge. Recently, Lin \textit{et al.} \cite{lin2022metaverse} provide an overview of Metaverse in education; 2) Entertainment. It seems that Metaverse in the entertainment market (e.g., playing games, watching a movie, and singing) is most successful currently. The digital reality experiences solve the limitations of space and time. Meanwhile, recreational activities can greatly assist students in broadening their interests; and 3) Security and privacy. It is doubtless that Web 3.0 architecture plays a vital role in data protection. That is, the user can control all aspects of their usage data. Decentralized identity effectively prevents users from identity theft and many cybercrime \cite{chen2022metaverse}. For lack of space, we do not list all aspects of Metaverse influences. In brief, the Metaverse will have a disruptive impact on smart cities, social activity, and economics within Web 3.0 \cite{sun2022metaverse,sun2022big,abrol2022web}.

Based on previous experience in the development of the Web, Tony Parisi proposed ``The Seven Rules of the Metaverse'' \cite{parisi2021the}. That is, the Metaverse should be unique (rule \#1) and can consistently be self-upgraded (rule \#7); the Metaverse should serve everyone (rule \#2) and be open to everyone (rule \#4); the Metaverse is not controlled by anyone (rule \#3); the Metaverse plays as an accessible network (rule \#6) and it should be hardware independent (rule \#5). Though the Metaverse is suspected of being overhyped, this does not prevent it from providing a more reliable environment for humankind in the future. Metaverse is just a tool for users to experience the virtual world. The immersive experience greatly improves the ability of users' sensory perception. Metaverse will also forever change the internet devices users adopt. The relationship between Metaverse and Web 3.0 is more like the productive force and the relations of production. The Metaverse will eventually infiltrate every aspect of our lives. The productive force that Metaverse brings demands for novel relations of production. Web 3.0 ensures that the relations between productions will retain decentralization, data ownership, trustlessness, intelligence, connectivity, and ubiquity features. Web 3.0 offers basic web technologies and economic support for the Metaverse, which will facilitate the booming of productive forces. In return, the development of Metaverse will continuously promote the maturity of Web 3.0. The decentralized relationship of productivity, for example, is a catalyst for creator economy reform, and the new economic model allows human work to better reflect labor value than previously.

\subsection{AI Generated Content}

Recently, ChatGPT\footnote{\url{https://openai.com/blog/chatgpt/}} has attracted the most attention within academia and industry fields. As an AI chatbot, ChatGPT displays excellent understanding and the ability to write. The most impressive thing is that ChatGPT supports multiple rounds of conversations and responses in real time. From OpenAI's first Generative Pre-trained Transformer (GPT) model to GPT-3 \cite{brown2020language}, to instructGPT \cite{ouyang2022training} and then ChatGPT, the iterative evolution of the model has brought many surprises to people. AI has shown its remarkable ability in the field of Natural Language Processing (NLP). In recent years, humans have been the great creative force in art, literature, science, and technology. Nonetheless, AI-Generated Content (AIGC) is gaining popularity as a new mode of content production on the internet. It is foreseeable that NLP technology represented by chatGPT will undoubtedly be adopted in massive application scenarios in the future, such as no-coding programming, novel generation, conversation search engines, voice companions, artificial intelligence customer service, and machine translation. Besides, an AI-generated picture surprisingly won the blue ribbon in the fair's contest for emerging digital artists \cite{roose2022ai}. Though it brings some worries about ethical concerns and joblessness, we suppose it is a chance to reconsider art itself. When the camera first took a photograph, most people did not expect the birth of photorealism.

AI is not only a reliable and smart assistant for humans but also a productive generator within Web 3.0, which will enrich the internet world. Due to the fact that AI is better than humans at mining knowledge and organizing material, AIGC has revealed its great potential in creation. AIGC will produce surprisingly good results if we provide enough data. Moreover, AIGC technology can help the digital human be smarter because AIGC is able to enhance the digital human's language understanding, action interaction, and emotion expressiveness. Similarly, Web 3.0 is able to advance in the field of digital content. The Al tools help us solve any video or image labeling task 10x faster and with 10x less manual work than before. In fact, the cultural treasures of human history are priceless because they are the result of human ingenuity. Humanity has evolved over more than six million years, and AI cannot rival our creative abilities forever. Although the purpose of Web3 is to protect the data ownership of creators from monopolies, most current online platforms are increasingly focusing on driving private traffic to earn revenue, which splits the internet world. The AIGC prevents this wrong trend, such as ChatGPT. The open API culture should be respected and continuing. In the Web 3.0 era, AIGC will be a common tool to assist people in content creation, which is cost-effective and greatly improves the quality of content.

\section{Challenges and Issues} \label{sec:challenges}

Although Web 3.0 will finally break the data monopoly of centralized enterprises, there are a lot of challenges and issues that should be carefully considered and solved. All of the above topics can be roughly classified into social, financial, legal, and technological categories. We mainly discuss the top three parts below, and the last one we have already discussed before.

\subsection{Socialization}

Nowadays, most people feel intolerable if their ties with the outside world are cut. We may have a strong desire to connect with others, whether they are nearby or at the other end of computers or phones. Cyberspace provides a good place for us to communicate with each other, and we express our thoughts and emotions freely there. In most cases, people surfing the Internet are communicating with people (whether known or unknown) who are not around, and we always expect to obtain reactions from others by sharing our own experiences. However, in general, the reality is not as good as you imagine it to be. Socializing in cyberspace may meet more troubling issues than offline, such as the dissemination of rumors, racial discrimination, terrorism, and negative impacts on youth. Despite the fact that Web 3.0 is supposed to comprehend human expressions and respond appropriately, we have to realize that there is still a long way to go before achieving this goal. As shown in Figure \ref{tab:products_web_3.0}, though Steemit uses rewards to encourage users to create or find valuable posts, the group behavior is troubling. Web 3.0 still requires further research to solve these problems.

\subsection{Independence}

Can Web 3.0 completely abandon the Web 2.0 architecture? The answer that we believe is no. The prototypes of social media and networks already exist. As we mentioned, Web 2.0 has greatly enriched people's lives on the Internet. It covers most users' online activities and causes path dependence for certain companies. Notice that it is hard for users to suddenly change their habits and accept new products. We take Steemit again as an example. Most of its functions are essentially the same as those of Twitter. In the view of users, both decide whether to push a post through others' likes, and others can forward their favorite posts to promote them. The significant difference is that Steemit will reward users, while Twitter may just increase a few users' followers. In addition, many functions of existing Web 3.0 applications are not perfect and still require the support of the Web 2.0 architecture (e.g., browsers and regulations). In general, considering the current development of Web 3.0 technologies, we cannot completely get rid of the influence of Web 2.0. We believe that Web 2.0 and Web 3.0 will coexist or even be complementary for a long period of time.

\subsection{Finance and Crime}

Up to now, there are many decentralized finance products (e.g., NFT, cryptocurrency, and crypto exchange) in usage. These products all claim that they can or will break the constraints of the original centralized value-exchange financial system. Through decentralization, the new financial services are more open, transparent, and interoperable. However, we hold several viewpoints about the new financial system. The most-running Web3 products\footnote{Please attention! Not Web 3.0 products.} today are based on their own special cryptocurrencies. Many transactions are done through these tokens (i.e., cryptocurrencies). Certainly, all deals between users are verifiable with the help of blockchain and smart contracts. What's more, the key premise on which these transactions normally proceed is the credibility of the token. In other words, the value of a token depends on how well it is accepted. Unfortunately, the recent bankruptcy of the FTX exchange event \cite{kevinis2022} and Luna coin \cite{ryanthe2022} shows that the cryptocurrency is unreliable\footnote{Crypto exchange is an online financial platform that allows buyers and sellers to trade cryptocurrencies. There are decentralization and centralization models.}. Since smart contracts provide users with many vital services, such as hosting and transaction processing, the decentralization exchange totally relies on the security of smart contracts. Until now, most decentralization exchanges hired or outsourced employees to complete code auditing work. It is not only more difficult but also more expensive. 

\subsection{Governance and Organization}

As the saying goes, there are two sides to a coin. Web 3.0 can solve data ownership issues and promotes data protection, while it also brings some challenges in governance, such as a regulatory puzzle, heavier wealth gaps, and money laundering \cite{turi2020currency}. It seems that someone who holds numerous cryptocurrencies will be the new monopoly in Web 3.0. The old monopolies have the first-mover advantage because of their wealth. Due to the data protection mechanism, it is hard for the public to regulate the profitable ways of corporate monopolies. The government faces more difficult troubles, such as cracking down on economic crime. For instance, someone who engages in financial fraud, stock manipulation, or insider trading will be punished because a centralized government must maintain the fairness of trading and prevent illegal activities. However, the slogan ``To the moon'' can let Dogecoin generate a return of about 150 times, and the ``hustle'' word means that the token has lost almost 92\% of its value since. This is a disguised plunder of wealth, but no one was punished for it. This causes instability in society \cite{kiayias2022sok}. As a result, a series of unresolved arguments about data, surveillance, competition, and security will spill over from the virtual world into the real one. Besides, it is worth noting that cryptocurrency is not equal to the whole economic construction of Web 3.0. Governments should enact the necessary regulatory standards in order to assist Web 3.0 in safely and smoothly completing its brutal stage.

\subsection{Law-making}

Additionally, the emergence of cryptocurrencies, particularly Bitcoin, has shaken governments' control over the traditional financial system and currency issuance. In order to maintain the sovereignty of the country's currency control, countries take different actions to regulate decentralized finance activities. For example, in China, government legislation prohibits cryptocurrency trading, but the country charges ahead with its digital yuan (abbreviated as e-CNY)\footnote{\url{https://en.wikipedia.org/wiki/Digital_renminbi}}, which is equal to the legal tender. The American government created the Digital Dollar Project\footnote{\url{https://digitaldollarproject.org}} has given rise to extensive research and discussion. The project aims to explore solutions in cyber resilience, financial inclusion, and other key areas for the next century. In conclusion, current legislation and governance mechanisms are still inadequate. The advent of a true Web 3.0 seems like a distant aspiration. There are many blanks that should be filled in the novel governance system, and we suppose that sufficient discussion and research should be made on how to ensure fairness and reliability within Web 3.0 architecture.

\section{Web3 vs Web 3.0} \label{sec:discussion}

Because Web 3.0 and Web3 are both extensions of the Semantic Web, there is a common misconception that they are interchangeable. In fact, as with the opinion of Timothy John Berners-Lee, there are relatively great differences between Web 3.0 and Web3. Web 3.0 is a web-based on smart input terminals. The distributed system is largely its core idea and then implements a decentralized network, but Web3 is more about decentralized governance with blockchain technology. Then, Web3 incorporates some economic elements, e.g., non-fungible tokens \cite{wang2021non}, an incentive model, and value exchange. Hence, its state of commercialization is higher than that of Web 3.0. In contrast, Web 3.0 is more of an academic topic than a commercial project. It is undeniable that the profit-driven model can significantly accelerate the development of new things in most cases. However, things are not always to our satisfaction. For example, most people have high hopes for non-fungible tokens because they can ascertain the property rights of digital assets. Indeed, most current non-fungible token business projects just pursue the possibility that they will make their investors and creators rich but ignore the utility of what is being created \cite{o2021too}.

In addition, Web3 puts more emphasis on trust as a dependency. In other words, the Web3 architecture needs one or many stable reputation systems to ensure its reliability \cite{keizer2021case}. However, in the case of the Luna coin \cite{ryanthe2022}, its currency value peaked at \$119.5 per token and then dropped as low as \$0.12 per token. During those three years, there was such a significant wake! This case illustrates that the credibility of cryptocurrency is not controllable, and the unknown risk is the most dangerous. Web 3.0 is largely based on mature distributed technologies. It incorporates these technologies into decentralized thinking and ensures that data ownership belongs to the users themselves. Due to the fact that all information interactions rely on the main chain to complete data dissemination, the correctness, and stability of the main chain will directly influence the safety of Web3. Web 3.0 utilizes centralized, decentralized, or distributed networks to form a communication network, and third parties have to request data usage permission because data belongs to users. In conclusion, Web 3.0 shows better compatibility with the current Web architecture than that of Web3.

\section{Conclusion} \label{sec:conclusion}

As people's imagination of the next generation of the web, Web 3.0 is always full of disputes and disagreements and is supplemented by people in different periods. From the original semantic Web to the current decentralized web, with the continuous iteration of technologies and concepts, the Internet of Everything has become more intelligent, 3D, and decentralized, which are becoming the prominent labels of Web 3.0. However, the gift of decentralization can easily morph into a curse. Web 3.0 stresses decentralization and delegates power to users. This is a good willingness, but there may also be some regulatory problems. Reviewing or prosecuting hate speech, violence, and terrorism might be more difficult because of the decentralization of power. Furthermore, the development of Web 3.0 is still in its early stages. This means that technological innovation, the implementation process, and the associated risks are still evolving. In view of this, it seems that Web 2.0 and Web 3.0 will be co-existing for a long time.

This article provides an overview of how Web 3.0 will affect our daily and future lives. There may be some technologies or ideas that we haven't discussed in this article, even if we have tried our best to review related literature. We hope that this survey will help to identify potential research directions while investigating and studying Web 3.0. Massive studies and cases have shown that combining with Web 3.0 is a viable way to achieve relative equality in the virtual world. Novel technologies break down many barriers (such as data ownership, cost, and limited experience) that are difficult to solve in real life. Web 3.0 provides excellent visualization that is not available in Web 2.0. More research works (e.g., decentralized storage, edge computing, artificial intelligence, and socially linked data protocols) are required for further study due to the rapid development of technology. Besides, it is worth noting that the paper also draws attention to new ethical and criminal issues. How does Web 3.0 solve Web 2.0 problems? What new things will Web 3.0 bring? These issues are briefly discussed.

\begin{acks}
    This research was supported in part by the Fundamental Research Funds for the Central Universities of Jinan University (No. 21622416), Guangzhou Basic and Applied Basic Research Foundation (No. 202102020277), National Natural Science Foundation of China (Nos. 62002136 and 62272196), Natural Science Foundation of Guangdong Province (No. 2022A1515011861), the Young Scholar Program of Pazhou Lab (No. PZL2021KF0023), and Guangdong Key Laboratory for Data Security and Privacy Preserving.
\end{acks}

% \newpage

\bibliographystyle{ACM-Reference-Format}
\balance
\bibliography{paper}

\appendix

\renewcommand\thefigure{\Alph{section}\arabic{figure}}
\renewcommand\thetable{\Alph{section}\arabic{table}}
\section{Appendix} \label{sec:appendix}

\setcounter{figure}{0}
\begin{figure}[H]
    \centering
    \includegraphics[trim=0 0 0 0,clip,scale=0.25]{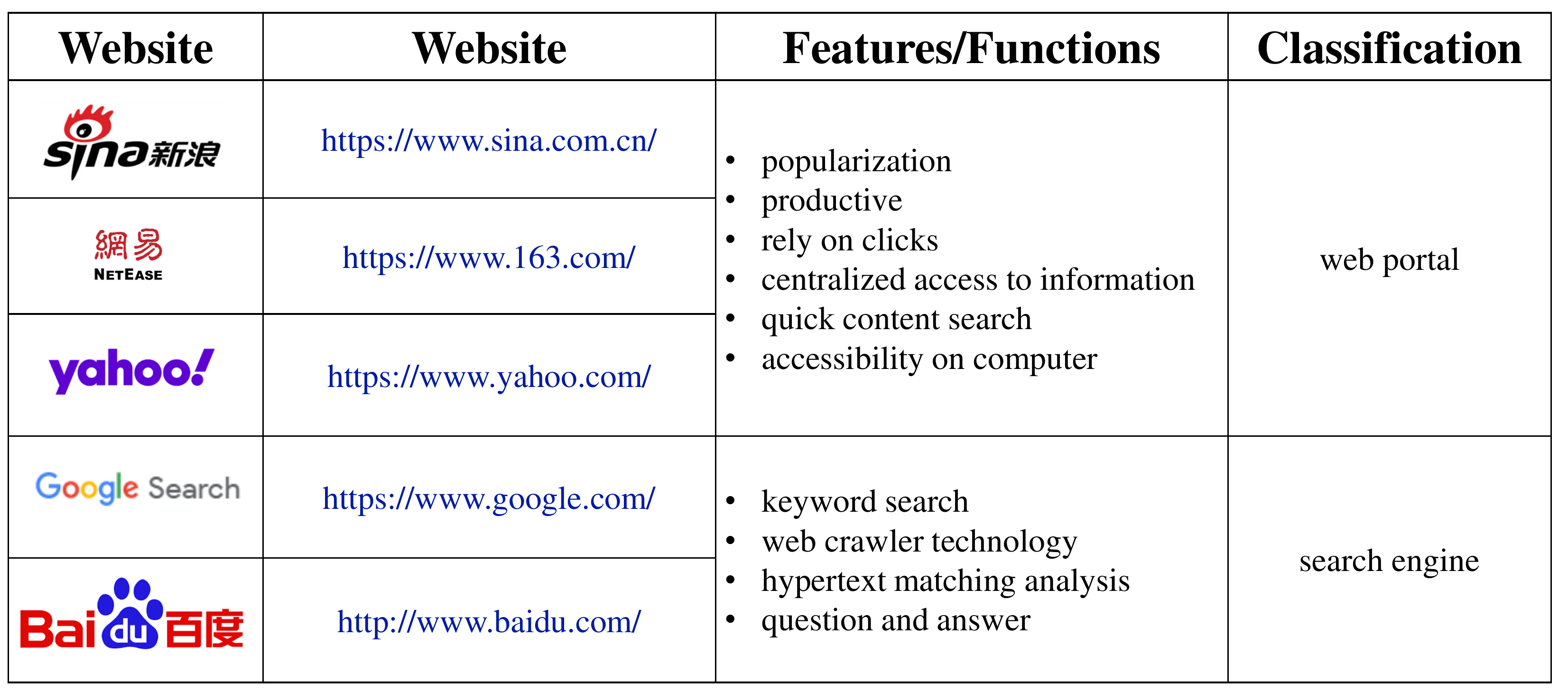}
    \caption{Several representative products in Web 1.0.}
    \label{tab:products_web_1.0}
\end{figure}

\setcounter{figure}{1}
\begin{figure}[H]
    \centering
    \includegraphics[trim=0 0 0 0,clip,scale=0.25]{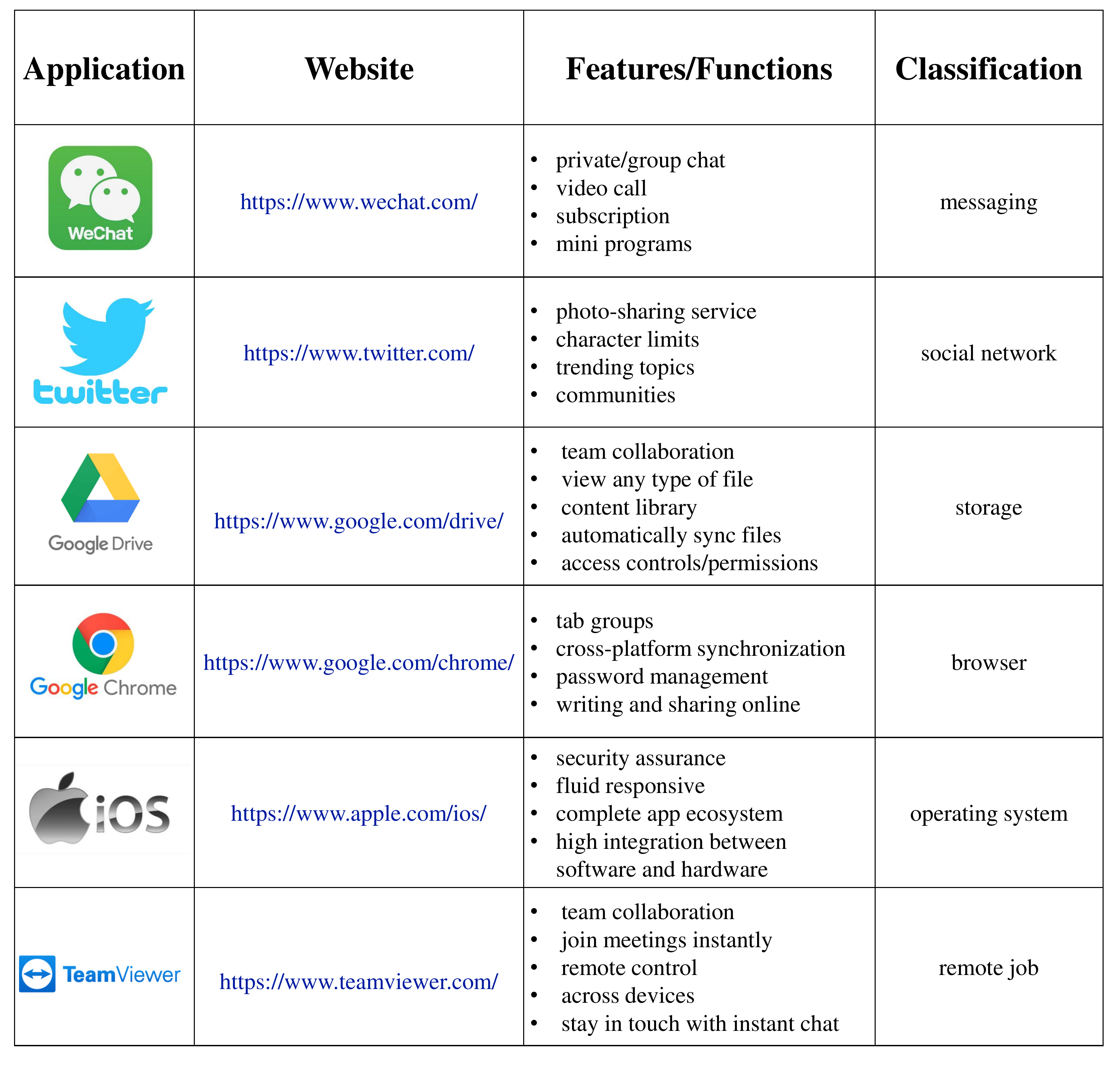}
    \caption{Several representative products in Web 2.0.}
    \label{tab:products_web_2.0}
\end{figure}

\setcounter{figure}{2}
\begin{figure}[H]
    \centering
    \includegraphics[trim=0 0 0 0,clip,scale=0.25]{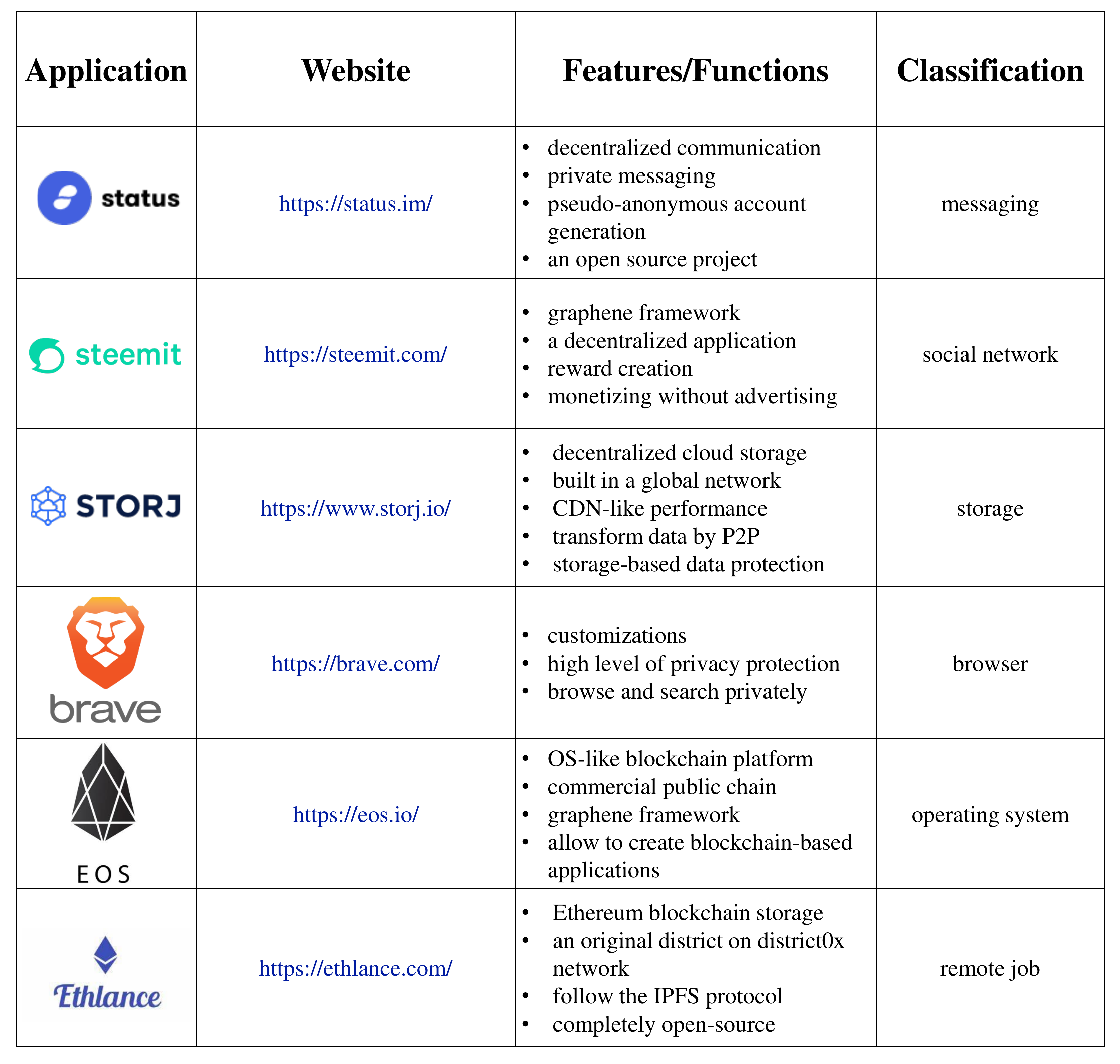}
    \caption{Several representative products in Web 3.0 or Web3.}
    \label{tab:products_web_3.0}
\end{figure}

\end{document}